\documentclass[9pt,twocolumn,twoside]{osajnl}

\journal{jocn} 

\setboolean{shortarticle}{false}
\title{Multi-Wavelength Transponders for High-capacity Optical Networks: A Physical-layer-aware Network Planning Study}

\author[1,2,*]{Jasper Müller}
\author[3]{Ognjen Jovanovic}
\author[1]{Tobias Fehenberger}
\author[1]{Gabriele Di Rosa}
\author[1]{Jörg-Peter Elbers}
\author[2]{Carmen Mas-Machuca}

\affil[1]{ADVA, Fraunhoferstr. 9a, 82152 Martinsried/Munich, Germany}
\affil[2]{Chair of Communication Networks, Technical University of Munich, Arcisstr. 21, Munich, Germany}
\affil[3]{DTU Electro, Technical University of Denmark, 2800 Kgs. Lyngby, Denmark}

\affil[*]{Corresponding author: jmueller@adva.com}

\doi{\url{http://dx.doi.org/10.1364/JOCN.483320}}

\usepackage{tikz}
\newcommand\copyrighttext{%
  \footnotesize \textcopyright~\the\year~Optica Publishing Group. One print or electronic copy may be made for personal use only. Systematic reproduction and distribution, duplication of any material in this paper for a fee or for commercial purposes, or modifications of the content of this paper are prohibited.
  \href{https://doi.org/10.1364/JOCN.483320}{DOI: 10.1364/JOCN.483320}.
  }
\newcommand\copyrightnotice{%
\begin{tikzpicture}[remember picture,overlay]
\node[anchor=north,yshift=-40pt] at (current page.north) {\fbox{\parbox{\dimexpr\textwidth-\fboxsep-\fboxrule\relax}{\copyrighttext}}};
\end{tikzpicture}
\vspace{-0.3cm}
}
\usepackage{hyperref}
\usepackage{color}
\usepackage{xcolor}
\usepackage{graphicx}
\usepackage{tikzscale}
\usepackage{tikz}
\usepackage{pgfplots}
\usepackage{caption}
\usepackage{subcaption}
\usepackage{verbatim}
\usepackage{multirow}
\usepackage{diagbox}
\usepackage{siunitx}
\usetikzlibrary{plotmarks,matrix,chains,scopes,fit,calc,shapes,positioning,decorations,intersections,fit,backgrounds,patterns}
\usetikzlibrary{fadings,shapes.arrows,shadows}
\pgfplotsset{compat=newest} 
\usepackage{forest}
\usepgfplotslibrary{groupplots}

\pgfplotsset{plot coordinates/math parser=false}
\pgfplotsset{every axis plot/.append style={solid,line width=1.5pt,mark size=1.5pt,mark options={solid,fill=white}}}
\pgfplotsset{every axis legend/.append style={legend cell align=left,font=\footnotesize}}
\newlength\FigureWidth
\newlength\FigureHeight
\newlength\FullFigureWidth
\graphicspath{{figures/}}

\pgfplotsset{myLegend/.append style={legend style={font=\footnotesize,at={(0.5,0.98)},anchor=north,align=left,legend columns=3}}}

\usepackage{xparse}
\tikzfading[name=arrowfading, top color=transparent!0, bottom color=transparent!95]
\tikzset{arrowfill/.style={#1,general shadow={fill=black, shadow yshift=-0.8ex, path fading=arrowfading}}}
\tikzset{arrowstyle/.style n args={3}{draw=#2,arrowfill={#3}, single arrow,minimum height=#1, single arrow,
single arrow head extend=.3cm,}}

\NewDocumentCommand{\tikzfancyarrow}{O{2cm} O{FireBrick} O{top color=OrangeRed!20, bottom color=Red} m}{
\tikz[baseline=-0.5ex]\node [arrowstyle={#1}{#2}{#3}] {#4};
} % https://tex.stackexchange.com/questions/84143/fancy-arrows-with-tikz

\makeatletter
\tikzset{
    block filldraw/.style={% only the fill and draw styles
        draw, fill=yellow!20},
    block rect/.style={% fill, draw + rectangle (without measurements)
        block filldraw, rectangle},
    block/.style={% fill, draw, rectangle + minimum measurements
        block rect, minimum height=0.8cm, minimum width=6em},
    from/.style args={#1 to #2}{% without transformations
        above right={0cm of #1},% needs positioning library
        /utils/exec=\pgfpointdiff
            {\tikz@scan@one@point\pgfutil@firstofone(#1)\relax}
            {\tikz@scan@one@point\pgfutil@firstofone(#2)\relax},
        minimum width/.expanded=\the\pgf@x,
        minimum height/.expanded=\the\pgf@y}}
\makeatother

\newcommand{\SNR}{\ensuremath{\text{SNR}}}
\renewcommand\footnotemark{}

\newcommand{\TXOSNR}{OSNR\textsubscript{TX}}
\newcommand{\Pline}{P\textsubscript{line}}

\newcommand{\rev}[1]{#1}

\dates{Accepted March 8, 2023}

\begin{abstract}
Continued cost- and power-efficient capacity scaling in optical networks is imperative to keep pace with ever-increasing traffic demands. In this paper, we investigate multi-wavelength transponders as a potential way forward. Suitable system architectures and realistic specifications of multi-wavelength transponders are identified and analyzed in terms of transmit OSNR penalties and spectral constraints. We investigate the performance for different specifications as compared to single-wavelength transponders in a network planning study on two network topologies, developing guidelines for multi-wavelength transponders specifications and their potential benefits. The studies show a reduction in the number of required lasers of up to 83~\% at the expense of a slight increase in number of lightpaths, demonstrating the potential for significant cost savings and efficiency improvements.
\end{abstract}

\setboolean{displaycopyright}{true}

\begin{document}

\maketitle
\copyrightnotice

\section{Introduction}
%% optics need big & efficient pipes, but the growth is not possible with per-lambda symbol rate any more

Optical networks build the backbone of today's telecommunication networks enabling bandwidth-hungry applications. New innovations such as high-resolution video streaming, 5G and autonomous driving lead to constantly increasing traffic demands in optical networks. Therefore, cost- and power-efficient solutions to capacity scaling are imperative in order to support the traffic growth. The increase of symbol rate per wavelength has been the main driver of decreasing cost per bit as well as power consumption, building up to recently announced coherent optical transceivers that support symbol rates of up to 140~GBd \cite{Jannu}.
These large bandwidths come with highly challenging requirements, in particular on the electronics. Therefore, it is uncertain for how much longer such a scaling of the symbol rate remains technically feasible and economically sensible.

%% we need optical multiplexing
A potential solution for continued cost-efficient capacity scaling is to deploy optical carrier multiplexing, i.e., to use several optical tributary signals on different wavelengths per transponder unit \cite{ITU}. Integrated multi-wavelength sources (MWSs) are using a single optical power supply, i.e., laser, such as an optical frequency comb to provide several lines in a single integrated component, thereby potentially offering significant efficiency improvements. Notable progress has been made in the MWS subsystem used to generate the lines~\cite{kuo2013wideband,gnauck2014comb,anandarajah2015enhanced,zhou201140nm,marin2017microresonator,imran_survey_2018}. MWSs have also been studied on a system~\cite{Pfeifle:15,schroder2019laser,Marin-Palomo:20} and architecture~\cite{sambo_sliceable_2014} level. 
MWSs have also been analyzed from a network perspective. Previous work covers novel routing, spectrum and transponder allocation algorithms~\cite{SantAnna} and an analysis of the impact of MWSs on the provisioning as well as the restoration of traffic~\cite{dallaglio_impact_2014}. Furthermore, different optical power supply options have been investigated with respect to techno-economic aspects~\cite{imran_techno-economic_2016} and a network throughput study has been performed~\cite{masood_smart_2020}. 
% However, 
\rev{Future high-baud rate transponders have also been investigated in a network planning context, identifying a trade-off between minimizing the number of transponders and spectral blockage in mesh networks \cite{Pedro:beyond100G}. Additionally, a design framework for the maximizing the spectral efficiency of flexible-grid optical networks has been proposed. A trade-off between unused reserved spectrum when planning for maximum spectral efficiency and a lower spectral efficiency when planning for just enough capacity is discussed \cite{Pedro:maxSE}. These trade-offs are also observed in network planning for MWSs.}
The impact of MWS specifications in a physical-layer aware network study has not yet been addressed. 
% It is important to evaluate MWSs and compare them to SWSs through network planning studies with respect to benefits and requirements. 
Quantifying the MWSs impact in different network scenarios will help to focus the development on specifications that are most beneficial as well as support network operators in choosing the most efficient solution for their network.
Therefore, we have analyzed the impact of MWS-based transponders on a network level as to provide guidelines to their specifications and required cost savings~\cite{ECOC_combs}.

%% what we do
In this work, we extend our previous comparison of \rev{single-wavelength source (SWS)} and MWS transponders~\cite{ECOC_combs}. For this study, the comparison is extended to 4-line and 8-line MWSs with fixed free spectral range (FSR), i.e., spacing of lines, as well as with flexible-FSR, in the following called fixed-FSR MWSs and flexible-FSR MWSs. We describe suitable forward-looking architectures for MWS transmitters and identify realistic transmit optical signal-to-noise ratio (\TXOSNR) values, depending on practical parameters. Using these values along with further constraints of MWSs, a network planning study is conducted on two topologies of different characteristics. 
We show that in optical mesh core networks, fixed-FSR MWSs can be beneficial only when deployed alongside SWSs. We analyze hybrid deployment strategies using the number of required lightpaths for a traffic demand as indicator on whether to use an MWS.
\rev{We study the impact of different numbers of required lightpaths as threshold for the selection of an MWS over multiple SWSs. We observe that higher thresholds for the deployment of MWSs lead to increased capacity as deploying less MWSs increases the flexibility in spectrum allocation. On the other hand, choosing higher thresholds decreases the potential cost-benefit of the deployment strategy.}
% We observe that choosing a higher threshold for the deployment of MWSs increases the traffic that can be provisioned without underprovisioning while potential cost-benefits decrease. 
\rev{For flexible-FSR MWSs, we observe a significant reduction in the number of required wavelength sources of about 70~\% for 4-line MWSs and about 83~\% for 8-line MWSs at the cost of only a moderate increase in the number of required lightpaths compared to SWS deployment. This is due to the \TXOSNR~penalty of MWSs. These results promise the potential of significant cost savings and efficiency improvements over SWSs. We show that the cost of the MWS block can reach up to 610~\% of the cost of an SWS to be economically viable. The presented study gives guidelines on the required MWS specifications and required savings in order for MWSs to become a viable alternative to SWS transponders in future efficient optical networks.}

\section{Transmitter architectures and properties}\label{sec:architectures}
In current transmitter architectures, an array of SWSs is used and each is used for a wavelength-division multiplexed (WDM) channel. Typically, a laser characterized by an optical carrier-to-noise ratio (OCNR) and output power $\text{P}_{\text{out}}$ is used as an SWS. The optical signal is modulated by an I/Q modulator and it is combined with the other channels by a multiplexer (MUX) to form a WDM signal. The combined signal is amplified by a booster amplifier (BA) to achieve a launch power per channel of $0$~dBm \rev{as a common value for a typical scenario of a 32~GBaud channel in a standard single-mode fiber network}. In this work, we consider an SWS with OCNR=$55$~dB and $\text{P}_{\text{out}}$=16~dBm. \rev{The modulator insertion loss and other sources of signal attenuation in the transmitter, such as laser power splitter, polarization beam splitter and polarization beam combiner, add up to 23~dB~\cite{Loss_ref}. Additionally, a MUX with a loss of~5~dB~\cite{Loss_ref} is assumed, as well as a 5~dB modulation loss independent of the QAM format~\cite{Marin-Palomo:20}.} An erbium-doped fiber amplifier (EDFA) with a noise figure of 5~dB is used as the BA, resulting in the generation of amplified spontaneous emission (ASE) noise with power \cite{Marin-Palomo:20}:

\begin{equation}
    P_n= p h f_c B_{ref} \frac{F_n(G-1)}{2}, 
     \label{eq:1}
\end{equation}
where $p$ is the number of polarization states considered, $h$ is Planck's constant,  $f_c$ is the carrier frequency and $B_{ref}=12.5$~GHz is the reference bandwidth corresponding to $0.1$~nm. The gain $G$ of the BA is calculated to obtain the desired launch power per channel. 
% \rev{For the given parameters, the {\TXOSNR} of this architecture is around 36~dB.}
\rev{For $\text{P}_{\text{out}}$=16~dBm and combined losses of 33 dB, a gain of 17 dB is required to achieve a launch power of 0~dBm per channel, resulting in a {\TXOSNR} of 36~dB for this architecture.}

\begin{figure}[!t]
\centering
  \includegraphics[width=\linewidth]{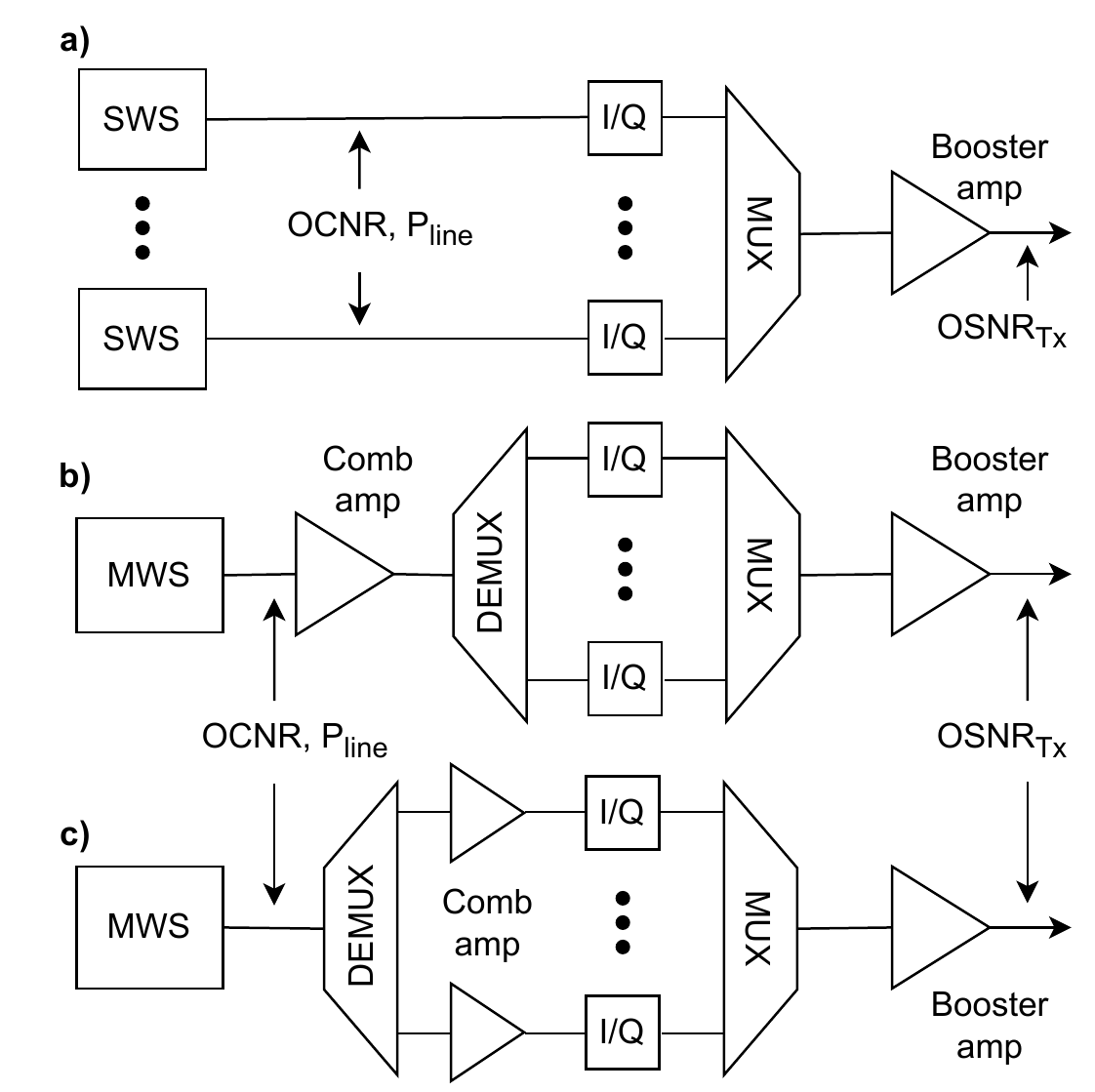}
 \caption{\rev{Transmitter architectures using SWSs \textbf{a)} and MWS} with \textbf{b)} joint amplification of all lines and \textbf{c)} per line amplification.}
 \label{fig:CombTxArch}
\end{figure}

A potential way to improve the cost-efficiency of transmitters based on arrays of SWSs is to replace them with integrated terminals exploiting a reduced number of MWSs. An MWS is an optical frequency comb that provides multiple equally spaced optical carriers from a single laser. Typically, an MWS is described on a system level by its free spectral range (FSR), number of lines, power per line (\Pline), and OCNR. The working principle of this device is to modulate an SWS to generate multiple coherent lines. However, this process comes at the cost of lower optical power and degraded OCNR per line. Transmitters require a sufficiently high power per line to perform I/Q modulation while satisfying the output signal power requirements, therefore amplification of the MWS may be mandatory for most practical applications. \rev{In Fig.~\ref{fig:CombTxArch}, the reference architecture using SWSs ((referred to as \textbf{a)}) as well as two potential architectures for MWSs are considered: amplification of all the lines with a single amplifier (referred to as \textbf{b)}) and individual amplification of each line (referred to as \textbf{c)}).} The former is reported in Fig.~\ref{fig:CombTxArch}~\textbf{b)}, where a comb amplifier (CA) follows the MWS to amplify all the lines jointly. The amplifed lines are then separated by a demultiplexer (DEMUX) to modulate each of them separately with an I/Q modulator and independent data streams. Afterward, a MUX combines the modulated lines into a single optical signal \rev{which is amplified by a BA to achieve the targeted launch power per channel.}
% a launch power per channel of $0$~dBm. 
The advantage of this architecture is that it requires only a single additional amplifier compared to an SWS array, however, the CA has a limited output power which is typically up to 26~dBm \cite{ThorLabsManual}. This bottleneck is overcome in the architecture shown in Fig.~\ref{fig:CombTxArch}~\textbf{c)} by using a CA for each line. Here, the lines are first separated by a DEMUX and then individually amplified by a CA before the modulation. In this way the amplifier output power is not a limiting factor, however, the number of required amplifiers increases linearly with the number of lines. We consider that the DEMUX has a loss of $L_{\text{DEMUX}}=5 \text{ dB}$~\cite{Loss_ref} and that the CA is an EDFA with a noise figure of 5~dB.

Due to the additional amplifiers and lower OCNR, both considered MWS architectures have a degraded {\TXOSNR} compared to the standard SWS, which is used as the reference in this work. The {\TXOSNR} of the MWS for $\text{N}_\text{lines}=4$ and $\text{N}_\text{lines}=8$ lines will be calculated for the two architectures. In both cases, the gain of a CA is calculated as
\begin{equation}
    \text{G}_\text{CA} = \text{P}_{\text{out}} - (\Pline - L_{\text{DEMUX}}),
\end{equation}
where the desired power per line that goes into the I/Q modulator is the same as for the SWS, i.e. $\text{P}_{\text{out}}=16$~dBm. However, as aforementioned, the output power of the amplifier is typically limited. Therefore, when considering architecture \textbf{a)}, the gain of the amplifier needs to be adjusted to avoid exceeding this physical limit. Assuming that all lines have the same \Pline, the output power of the amplifier can be calculated as
\begin{equation}
    \text{P}_\text{CA,out} = (\Pline + \text{G}_\text{CA}) + 10\log_{10}(\text{N}_\text{lines}).
\end{equation}
If this power exceeds the limit, the gain is decreased to preserve the constraint $\text{P}_\text{CA,out} \leq 26$~dBm. For architecture \textbf{b)}, the power per line after the amplifier is always 16~dBm. Finally, the {\TXOSNR} is calculated as
\begin{equation}
    \text{OSNR}_{\text{TX}}^{-1} = (\text{OCNR}^{-1} + \text{OSNR}_{\text{CA}}^{-1} + \text{OSNR}_{\text{BA}}^{-1}).
\end{equation}
In \cite{Marin-Palomo:20}, it was considered that the modulator acts as a polarizer and that only the noise co-polarized with the carrier is considered (Eq.~\ref{eq:1}). Therefore, we have considered that the noise contribution of CA is for a single polarization $p=1$, whereas the noise contribution of BA is for dual polarization $p=2$. It should be noted that the OCNR is also considered with respect to the co-polarized noise.
\rev{While we assume fixed parameters for the SWS transponder and therefore a fixed {\TXOSNR} of 36 dB, a decreased SWS {\TXOSNR} would lead to a lower influence of the additional CAs of MWS transponders while a higher SWS {\TXOSNR} would increase the influence. In a common window of 36 to 40 dB no notable difference would be observed in {\TXOSNR} penalty of MWS compared to SWS transponders.}
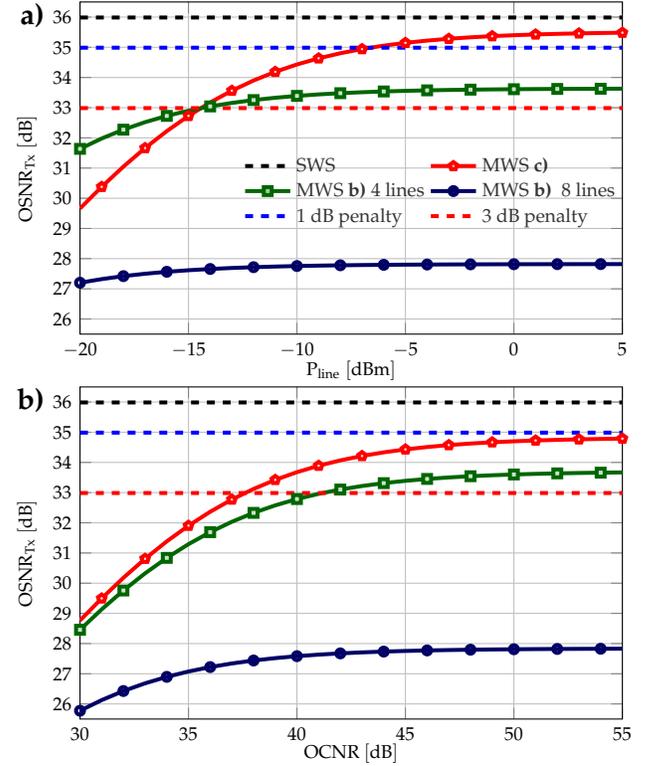
\begin{figure}[!t] % There was an error in the ECOC results, I have corrected it now
\centering
 % directly input tikz without tikzscale
%   \tikzsetnextfilename{#1}%
  \pgfplotsset{compat=newest}

\pgfplotsset{
    discard if not/.style 2 args={
        x filter/.code={
            \edef\tempa{\thisrow{#1}}
            \edef\tempb{#2}
            \ifx\tempa\tempb
            \else
                \def\pgfmathresult{inf}
            \fi
        }
    }
}

\begin{tikzpicture}[font=\footnotesize]
    \begin{groupplot}[group style={group size=1 by 2, vertical sep=20pt}]
    \nextgroupplot[
        width=\columnwidth,
        height=6cm,
        grid=both,
        title={\Large \textbf{a)}},
        title style={at={(-0.05,0.85)},
                   anchor=south east},
        ymax = 36.5, ymin = 25.5,
        xmin = -20, xmax = 5,
        % grid style={line width=.1pt, draw=gray!10},
        % major grid style={line width=.2pt,draw=gray!50},
        ytick distance=1,
        xtick distance=5,
        xlabel shift = -5,
        xlabel=\footnotesize $\text{P}_{\text{line}} \text{ } \si{[dBm]}$,
        ylabel=\footnotesize $\text{OSNR}_{\text{Tx}} \text{ } \si{[dB]}$,
        % legend pos = south east,
        legend style={font=\normalsize,nodes={scale=0.8, transform shape},fill opacity=0.8, at={(1.01,0.3)}, anchor=south east, draw=none, fill=none},
        legend columns=2]
        \addplot+[discard if not={method}{CW_laser},no marks,color=black,style=dashed] table [x=Pline, y=OSNR, col sep=comma] {Figures/OSNR_Tx_Pline.csv};    
        \addlegendentry{SWS}
        % \addplot+[discard if not={method}{Comb_arch1_no_cap},color=black, mark repeat=2,mark=*,mark options={fill=.!20!white}] table [x=Pline, y=OSNR, col sep=comma] {Figures/OSNR_Tx_Pline.csv};
        % \addlegendentry{MWS \textbf{a)} w/o cap}
        \addplot+[discard if not={method}{Comb_arch2},color=red, mark repeat=2,mark phase=2,mark=pentagon*,mark options={fill=.!20!white}] table [x=Pline, y=OSNR, col sep=comma] {Figures/OSNR_Tx_Pline.csv};
        \addlegendentry{MWS \textbf{c)}}
        \addplot+[discard if not={method}{Comb_arch1_w_cap},color=green!40!black, mark repeat=2,mark=square*,mark options={fill=.!20!white,solid}] table [x=Pline, y=OSNR, col sep=comma] {Figures/OSNR_Tx_Pline.csv};
        \addlegendentry{MWS \textbf{b)} 4 lines }%w/ cap}
        \addplot+[discard if not={method}{Comb_arch1_w_cap_8lines},color=blue!40!black, mark repeat=2,mark=o,mark options={fill=.!20!white,solid}] table [x=Pline, y=OSNR, col sep=comma] {Figures/OSNR_Tx_Pline.csv};
        \addlegendentry{MWS \textbf{b) } 8 lines }%w/ cap}
        \addplot+[discard if not={method}{Penalty_1dB},no marks,color=blue,style=dashed] table [x=Pline, y=OSNR, col sep=comma] {Figures/OSNR_Tx_Pline.csv};
        \addlegendentry{1 dB penalty}
        \addplot+[discard if not={method}{Penalty_3dB},no marks,color=red,style=dashed] table [x=Pline, y=OSNR, col sep=comma] {Figures/OSNR_Tx_Pline.csv};
        \addlegendentry{3 dB penalty}
        % \addplot+[discard if not={method}{Penalty_5dB},no marks,color=red,style=dashed] table [x=Pline, y=OSNR, col sep=comma] {Figures/OSNR_Tx_Pline.csv};
        % \addlegendentry{5 dB penalty}
    
    \nextgroupplot[
        width=\columnwidth,
        height=6cm,
        grid=both,
        ymax = 36.5, ymin = 25.5,
        xmax = 55, xmin = 30,
        title={\Large \textbf{b)}},
        title style={at={(-0.05,0.85)},
                   anchor=south east},
        %xmin = -20, xmax = 5,
        %grid style={line width=.1pt, draw=gray!10},
        %major grid style={line width=.2pt,draw=gray!50},
        ytick distance=1,
        xtick distance=5,
        xlabel=\footnotesize $\text{OCNR} \text{ } \si{[dB]}$,
        ylabel=\footnotesize $\text{OSNR}_{\text{Tx}} \text{ } \si{[dB]}$,
        xlabel shift = -5,
        legend pos = south east,
        legend style={nodes={scale=0.8, transform shape}},
        legend columns=2]
        
    \addplot+[discard if not={method}{CW_laser},no marks,color=black,style=dashed] table [x=OCNR, y=OSNR, col sep=comma] {Figures/OSNR_Tx_OCNR.csv};    
    %\addlegendentry{CW laser}
    % \addplot+[discard if not={method}{Comb_arch1_no_cap},color=black, mark repeat=2,mark=*,mark options={fill=.!20!white}] table [x=OCNR, y=OSNR, col sep=comma] {Figures/OSNR_Tx_OCNR.csv};
    %\addlegendentry{Comb arch1 no cap}
    \addplot+[discard if not={method}{Comb_arch1_w_cap},color=green!40!black, mark repeat=2,mark=square*,mark options={fill=.!20!white,solid}] table [x=OCNR, y=OSNR, col sep=comma] {Figures/OSNR_Tx_OCNR.csv};
    %\addlegendentry{Comb arch1 w cap}
    \addplot+[discard if not={method}{Comb_arch1_w_cap_8lines},color=blue!40!black, mark repeat=2,mark=o,mark options={fill=.!20!white,solid}] table [x=OCNR, y=OSNR, col sep=comma] {Figures/OSNR_Tx_OCNR.csv};
    %\addlegendentry{Comb arch1 w cap}
    \addplot+[discard if not={method}{Comb_arch2},color=red, mark repeat=2,mark phase=2,mark=pentagon*,mark options={fill=.!20!white}] table [x=OCNR, y=OSNR, col sep=comma] {Figures/OSNR_Tx_OCNR.csv};
    %\addlegendentry{Comb arch2}
    \addplot+[discard if not={method}{Penalty_1dB},no marks,color=blue,style=dashed] table [x=OCNR, y=OSNR, col sep=comma] {Figures/OSNR_Tx_OCNR.csv};
    %\addlegendentry{1 dB penalty}
    \addplot+[discard if not={method}{Penalty_3dB},no marks,color=red,style=dashed] table [x=OCNR, y=OSNR, col sep=comma] {Figures/OSNR_Tx_OCNR.csv};
    %\addlegendentry{3 dB penalty}
    % \addplot+[discard if not={method}{Penalty_5dB},no marks,color=red,style=dashed] table [x=OCNR, y=OSNR, col sep=comma] {Figures/OSNR_Tx_OCNR.csv};
    % \addlegendentry{5 dB penalty}
    %\end{axis}
    \end{groupplot}
\end{tikzpicture}

 \caption{{\TXOSNR} vs. power \textbf{a)} and OCNR \textbf{b)} per line for a MWS with four lines. The MWS curves relate to Fig.~\ref{fig:CombTxArch}.}%_{line}
 \label{fig:OSNR_OCNR}
\end{figure}

We now present the results concerning the {\TXOSNR} for the two MWS architectures of Fig.~\ref{fig:CombTxArch}. The {\TXOSNR} values obtained are shown in Fig.~\ref{fig:OSNR_OCNR} versus the power per line (\textbf{a)}) and OCNR (\textbf{b)}) for typical values of OCNR=45~dB and \Pline=$-\text{10}$~dBm, respectively \cite{HuOxenlowe}. We observe that to preserve less than 3~dB penalty in {\TXOSNR} compared to an SWS, the power (OCNR) per MWS line must be at least $-\text{14}$~dBm (40~dB), which is achievable by state-of-the-art MWSs \cite{Marin-Palomo:20,HuOxenlowe,schroder2019laser} when considering architecture \textbf{b)} with $\text{N}_\text{lines}=4$. However, if $\text{N}_\text{lines}=8$ is considered, a penalty of around 8~dB is observed. \rev{Due to this large penalty, we discard the choice of the \textbf{b)} architecture for 8 lines.} In the case of architecture \textbf{c)}, the {\TXOSNR} is independent of $\text{N}_\text{lines}$, however, at the expense of additional amplifiers, thus increasing the total cost of the terminal. 
% A trade-off between the two architectures is probably possible, but that was not the goal of this analysis.
% It should be mentioned that the degraded OCNR of the local oscillator is not considered in this study. 

% Need to explain that the second architecture is independent from the number of lines but it effects the cost because each line requires an amplifier. 

\begin{table}[ht!]
   \centering
\resizebox{\columnwidth}{!}{%
\begin{tabular}{|m{0.20\columnwidth}|m{0.13\columnwidth}|m{0.12\columnwidth}|m{0.18\columnwidth}|m{0.14\columnwidth}|m{0.15\columnwidth}|}
\hline
\textbf{Topology} & \textbf{\# Nodes} & \textbf{\# Links} & \textbf{\# Demands} & \textbf{Avg. Node Degree} & \textbf{Avg. Path Length} \\ \hline
Germany~\cite{sndlib}       & 17                & 26                & 136        & 3.05   & 420 km     \\ \hline
EU~\cite{sndlib}             & 28                & 41          & 378 & 2.92      & 1100 km        \\ \hline
\end{tabular}%
}
\caption{Core network topologies considered.}
\label{tab:table-topologies}
\end{table}

% \begin{table}
%  \caption{Transponder implementation penalties in dB.}
%  \label{tbl:configuration}
%  \centering
%  \begin{tabular}{|l||c|c|c|c|}
%  \hline
%  \diagbox[width=10em]{Modulation}{SR [GBd]} & 35 & 70 & 105 & 140 \\
%  \hline
%  QPSK  & 1 & 1.5 & 2 & 2.5  \\
%  \hline
%  16QAM  & 1.5 & 2 & 2.5 & 3 \\
%  \hline
%  64QAM & 2 & 2.5 & 3 & 3.5 \\
%  \hline
%  \end{tabular}
% \end{table}

\begin{figure}[t]
\centering
\includegraphics[width=\columnwidth]{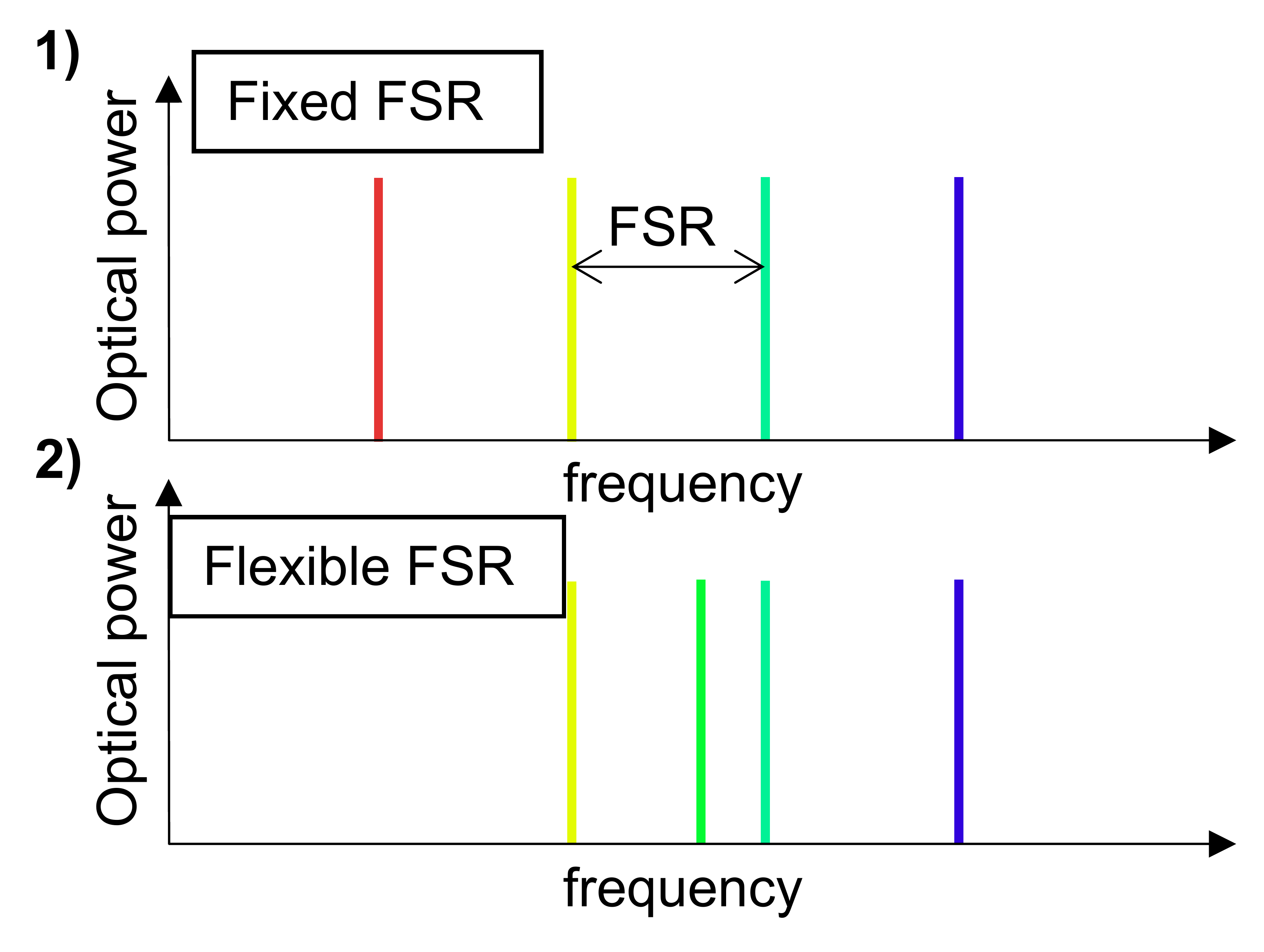}
\caption{Illustration of the spectrum of a 4-line fixed-FSR \textbf{1)} MWS and a 4-line flexible-FSR \textbf{2)} MWS.}
\label{fig:fixed_flex_comb}
\end{figure}

\section{Network Planning Study}
In this section, the implications of the use of MWSs in optical networks are investigated in an extensive network planning study. We compare in multiple planning scenarios the use of MWSs with different specifications to a baseline case in which SWSs are used exclusively. Two types of MWSs are considered, as illustrated in Fig.~\ref{fig:fixed_flex_comb}: an MWS with fixed-FSR (Fig.~\ref{fig:fixed_flex_comb} \textbf{1)}) limited to equal spacing between all lines and one with flexible-FSR (Fig.~\ref{fig:fixed_flex_comb} \textbf{2)}) capable of abitrary spacing between lines. For the flexible-FSR MWS, we assume no limitations in terms of center frequency, configuration and routing of the different lines compared to SWSs. In contrast, the fixed-FSR MWS is restricted to equal spacing between the lines. Additionally, all lines are required to co-propagate on the same route. However, the fixed-FSR is assumed to be adjustable to the channel spacing of any configuration considered for this study.

Our technical analysis aims at quantifying the aforementioned drawbacks imposed by MWSs regarding performance and flexibility, the latter being particularly critical for fixed-FSR MWSs. The results are then used to weigh these drawbacks against the cost savings enabled by the reduction in the number of wavelength sources. Since we investigate MWSs as a potential solution for capacity scaling in the future, forward-looking lightpath configurations of up to 140 GBd are considered in our study.

\begin{table}
 \centering
 \begin{tabular}{|l||c|c|c|c|}
 \hline
 \diagbox[width=10em]{Modulation}{SR [GBd]} & 35 & 70 & 105 & 140 \\
 \hline
 QPSK  & 6.2 & 6.7 & 7.2 & 7.7  \\
 \hline
 16QAM  & 13 & 13.5 & 14 & 14.5 \\
 \hline
 64QAM & 19.1 & 19.6 & 20.1 & 20.6 \\
 \hline
 \end{tabular}
  \caption{Required SNR in dB for the considered lightpath configurations.}
 \label{tab:configuration}
\end{table}

\subsection{Setup}
Two publicly available network topologies of different characteristics, representing a national (Nobel-Germany) and a continental (Nobel-EU) backbone network \cite{sndlib}, were chosen for this study. The main characteristics of these topologies are summarized in Table~\ref{tab:table-topologies}. For the planning scenarios, the links are assumed to be standard single-mode fiber (SSMF) links, consisting of 80~km spans with perfect attenuation compensation at the end of each span by an EDFA with 5~dB noise figure. We consider flexible WDM grid networks in the C-Band, with 400 frequency slots of 12.5~GHz each. The considered symbol rates (SRs) and modulation formats are listed in Table~\ref{tab:configuration}, together with the required SNR for each configuration. These values are obtained by adding modulations format and symbol rate dependent implementation penalties to the theoretical SNR that achieves the FEC threshold at a bit error ratio of 3.5\% for each QAM format. The transmit power spectral density is considered constant for all SRs. \rev{A pulse rolloff of 5\% is considered, leading to channel spacings of 37.5, 75, 112.5 and 150 GHz. In the MWS studies we consider the same channel spacing for comparison. For the fixed-FSR MWS study, it is assumed that the FSR can be chosen from the different channel spacings.}

\rev{In this study, we consider a traffic demand to be the aggregated traffic requests between a source-destination node pair.} 
The demands are weighted according to a traffic model based on the number of data centers and internet exchange points in each ROADM location \cite{patri2020planning}. To vary the network traffic demands, the individual demands are scaled by the same factor in order to reach different levels of aggregate requested network traffic (ART). The routing, configuration, and spectrum assignment (RCSA) algorithm considers $k=3$ shortest-path routing and uses the first-fit algorithm for spectrum assignment. The configurations are chosen in order to minimize the number of required lightpaths (LPs), with candidate configurations being chosen based on the linear SNR. In a second step, nonlinear interference is computed using the closed-form GN model~\cite{GN}. The SNR takes into account \TXOSNR, ASE noise from the amplifiers, and interference due to the fiber nonlinearities. Only configurations with a required SNR threshold lower than the computed SNR are considered. LPs that violate this condition are downgraded. When an LP is downgraded, the algorithm will try to activate additional lines available from a MWS or place additional LPs to fulfill the demand.

\subsection{Scenarios}
Different scenarios are investigated using the described network planning setup. First, planning using fixed-FSR MWSs is investigated. The goal of this study is to evaluate the feasibility of different deployment strategies of fixed-FSR MWSs in mesh core networks. We consider 4-line as well as 8-line MWSs and look at planning scenarios in which both fixed-FSR MWSs and SWSs coexist.
\rev{A trade-off between traffic that can be provisioned without underprovisioning and cost benefits has to be considered when selecting a hybrid deployment strategy. In this case, we consider simple strategies, where}
% For these scenarios,
we define $n_{\text{cutoff}}$ as the minimum number of LPs used to fulfill a demand between a source and destination pair when planning with transponders based on SWSs, to trigger the use of an MWS instead. This number is varied between 1 (equivalent to deploying only MWSs) and $\text{N}_\text{lines}$ (equivalent to deploying an MWS only if all lines are utilized). We will see in the following that this choice has a critical impact on the planning results.
In a second study, planning with conventional SWS transponders is compared to planning using flexible MWSs, assuming that the latter solution provides a \TXOSNR~penalty of 1~dB, 3~dB, or 5~dB. The goal is to translate this \TXOSNR~penalty of MWSs to the network planning to derive requirements on the devices.

\begin{figure*}[!t]
\centering
\includegraphics[width=\textwidth]{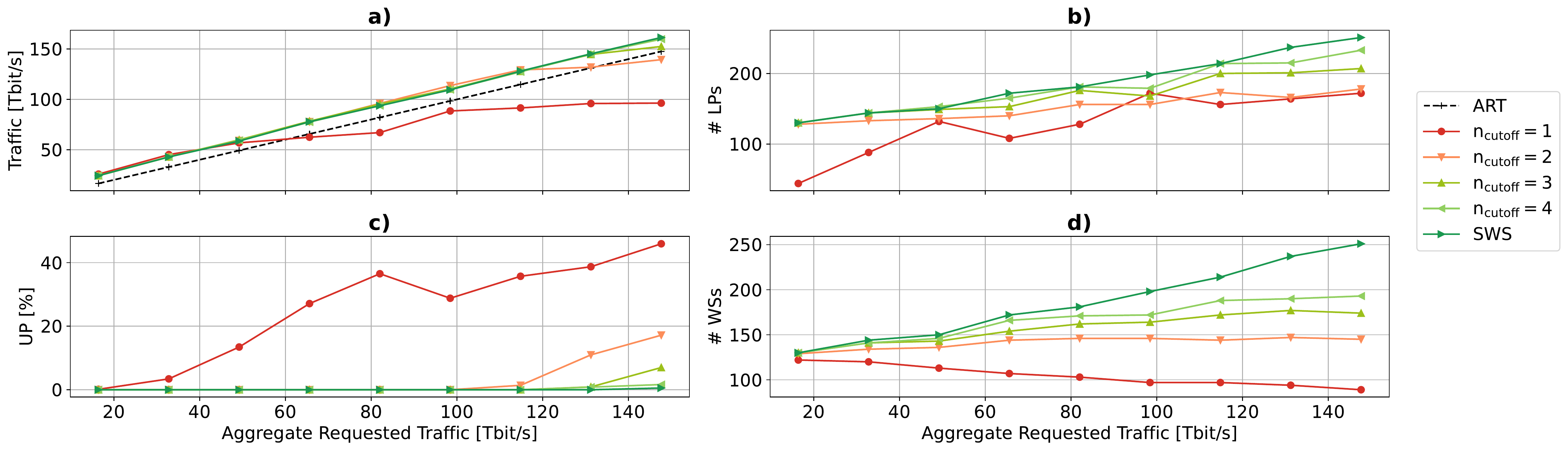}
\caption{Provisioned traffic \textbf{a)}, number of LPs \textbf{b)}, UP \textbf{c)} and number of wavelength sources (WSs) \textbf{d)} on Germany topology using fixed-FSR MWSs with 1~dB transmit OSNR penalty with different $n_{\text{cutoff}}$ compared to SWS baseline.}
\label{fig:fixed_study_4line_ger}
\end{figure*}

\begin{figure*}[!t]
\centering
\includegraphics[width=\textwidth]{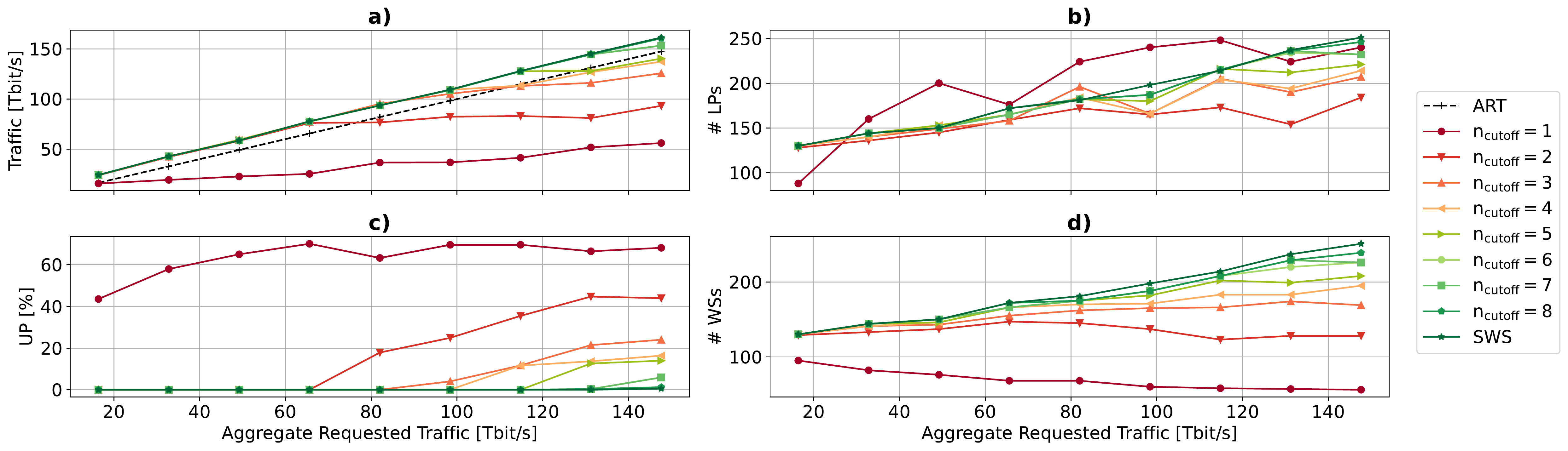}
\caption{Provisioned traffic \textbf{a)}, number of LPs \textbf{b)}, UP \textbf{c)} and number of WSs \textbf{d)} on Germany topology using fixed-FSR MWSs with 1~dB transmit OSNR penalty with different $n_{\text{cutoff}}$ compared to SWS baseline.}
\label{fig:fixed_study_8line_ger}
\end{figure*}

\section{Results and Discussions}
The network planning results are compared in terms of provisioned traffic, the number of deployed LPs and wavelength sources (WSs) for the different scenarios, and underprovisioning ratio (UP). UP is defined as \rev{the ratio of demanded traffic that cannot be provisioned over the overall demanded traffic }\cite{longterm_planning}
\begin{equation}
    \label{eq:up}
    UP =  \frac{\sum_{{\widetilde{d}}\in \widetilde{D}} \left ( DR_{\widetilde{d}} - \sum_{{lp} \in LP_{\widetilde{d}} }DR_{lp} \right )}{\sum_{{d}\in D} DR_d }.
\end{equation}
In \eqref{eq:up}, $DR_{d}$ is the requested traffic of demand $d$, and $DR_{lp}$ is the data rate of the lightpath $lp \in LP_{d}$ provisioned to carry traffic for demand $d$. $\widetilde{D}$ is the subset of demands where the lightpaths $lp\in LP_{\widetilde{d}}$ cannot satisfy the requested traffic and it is formally defined as:
\begin{equation}
    \label{eq:up_d_tilde}
    \widetilde{D} = \left \{ \widetilde{d} \in D \mid  DR_{\widetilde{d}} - \sum_{{lp} \in LP_{\widetilde{d}} }DR_{lp} > 0 \right \}.
\end{equation}
\rev{As the setup allows for partially provisioned demands, UP is the preferred metric over blocking probability.}
\subsection{Fixed-FSR MWS}
The use of fixed-FSR MWSs in mesh core networks with multiple ROADM nodes presents additional difficulties to the network planning compared to SWSs. Frequency slots have to be reserved for all lines when placing an MWS to guarantee that they can be activated later, even if not all lines are needed at the time of deployment. Furthermore, as co-propagation is required, each MWS can only serve a single demand. Due to these restrictions, it is not feasible to place exclusively MWSs in a network. Therefore, strategies based on the coexistence of SWSs and MWSs should be considered.
In the related network planning studies we therefore considered different values for $n_{\text{cutoff}}$ for both 4-line as well as 8-line fixed-FSR MWSs. Fig.~\ref{fig:fixed_study_4line_ger} shows the results for 4-line MWSs on the Germany topology. It can be seen that for $n_{\text{cutoff}}=1$, meaning that only MWSs are deployed, the traffic falls below the ART even for low values (ART > 20~Tbit/s) (Fig.~\ref{fig:fixed_study_4line_ger} \textbf{a)}). Moreover, due to the \TXOSNR~penalty, more LPs are deployed for low ART values to serve the same amount of traffic when compared to the SWS scenario, as visible from the results in Fig.~\ref{fig:fixed_study_4line_ger}. As the spectrum fills up quickly, due to reserved slots for all MWS lines, the \rev{increase of the number of LPs with traffic is lower for SWS} while UP increases with higher ART (Fig.~\ref{fig:fixed_study_4line_ger} \textbf{c)}). \rev{We observe that when only deploying MWSs ($n_{\text{cutoff}}=1$) the number of wavelength sources deployed during planning decreases with higher ART, as shown in Fig.~\ref{fig:fixed_study_4line_ger} \textbf{d)}.} The reason for this somewhat unexpected behavior is that with increased demands, higher-bandwidth configurations are chosen for the MWS and the spectrum of bottleneck links fills up with a lower number of MWSs. Higher values of  $n_{\text{cutoff}}$ lead to better performance in terms of provisioned traffic (Fig.~\ref{fig:fixed_study_4line_ger} \textbf{a)}) and higher ART that can be served without UP (Fig.~\ref{fig:fixed_study_4line_ger} \textbf{c)}), but also an increased number of deployed wavelength sources (Fig.~\ref{fig:fixed_study_4line_ger} \textbf{d)}). 

\begin{figure*}[!t]
\centering
\includegraphics[width=\textwidth]{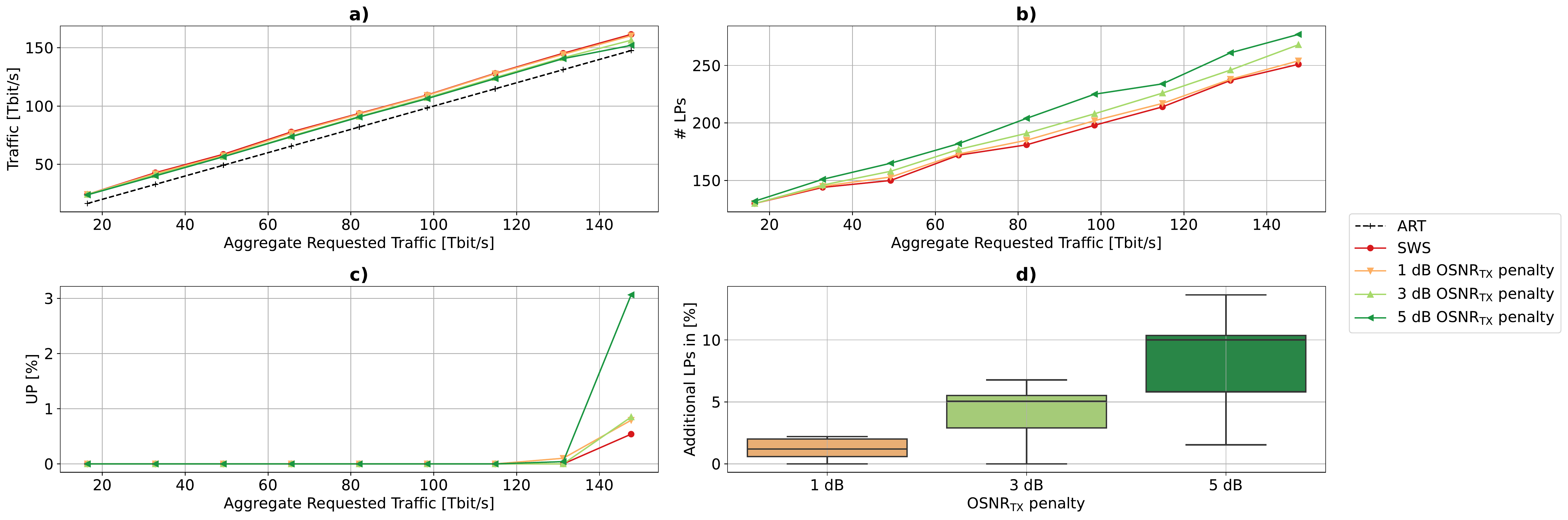}
\caption{Provisioned traffic \textbf{a)}, number of lightpaths \textbf{b)}, UP \textbf{c)} and distributions of additional wavelength sources \textbf{d)} on Germany topology using flexible-FSR MWSs with 1~dB / 3~dB / 5~dB transmit OSNR penalty compared to SWS baseline.}
\label{fig:flex_study_ger}
\end{figure*}

\begin{figure*}[!t]
\centering
\includegraphics[width=\textwidth]{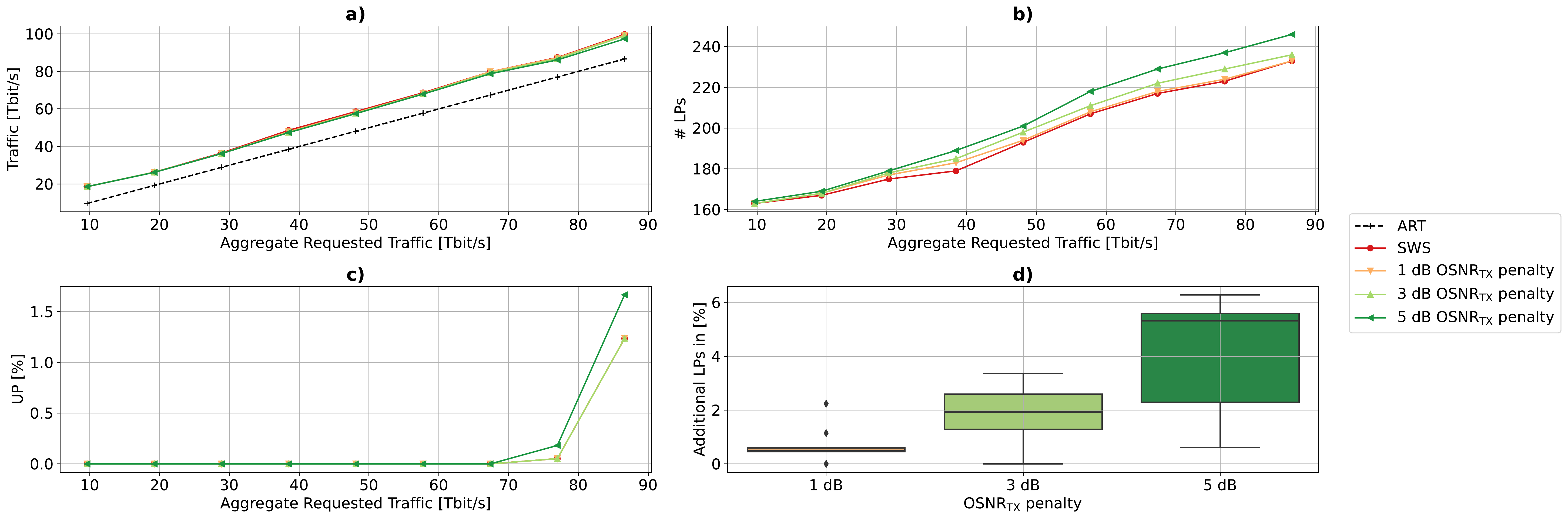}
\caption{Provisioned traffic \textbf{a)}, number of lightpaths \textbf{b)}, UP \textbf{c)} and distributions of additional wavelength sources \textbf{d)} on EU topology using flexible-FSR MWSs with 1~dB / 3~dB / 5~dB \TXOSNR~penalty compared to SWS baseline.}
\label{fig:flex_study_eu}
\end{figure*}

The results for 8-line MWSs on the Germany topology are consistent with the results obtained for the 4-line scenario. The strategy of only deploying MWSs leads to a low amount of provisioned traffic (Fig.~\ref{fig:fixed_study_8line_ger} \textbf{a)}), while the number of deployed LPs increases with the ART (Fig.~\ref{fig:fixed_study_8line_ger} \textbf{b)}). This is due to the sub-optimal nature of RCSAs. With higher demands, higher configurations are considered for the MWS lines. However, these configurations have to be downgraded in some cases due to the influence of nonlinear interference and \TXOSNR, leading to additional MWS lines that are activated to fulfill the demands. Therefore, while traffic and UP ((Fig.~\ref{fig:fixed_study_8line_ger} \textbf{c)}) stay at similar levels, the number of LPs increases and the number of wavelength sources decreases (Fig.~\ref{fig:fixed_study_8line_ger} \textbf{d)}) with higher ART. In this scenario, it can be seen that due to the limited bandwidth a UP of over 40~\% is observed even at the lowest considered ART. Therefore, the strategy of only deploying 8-line MWSs is not viable. As seen for the 4-line MWSs, the performance increases with higher $n_{\text{cutoff}}$, in parallel with a reduction in the savings concerning the number of wavelength sources required. For example, a solution that can fulfill 120~Tbit/s without underprovisioning ($n_{\text{cutoff}}\geq 3$ for 4 lines or $n_{\text{cutoff}}\geq5$ for 8 lines) deploys around 90~\% of the wavelength sources that would be needed using only standard SWSs.

We can conclude that fixed FSR MWSs should not be considered for general deployment in networks but can be useful and cost-effective tools when deployed for large demands that require multiple LPs. A trade-off between capacity and savings in the number of wavelength sources has to be found and a deployment strategy as well as the number of lines for the MWS should be chosen on a case-by-case basis depending on the network topology, traffic demands, and capacity requirements. The results for fixed-FSR MWSs on the EU topology are consistent with the results on Germany, showing only small differences for higher $n_{\text{cutoff}}$ and small benefits in terms of savings of the number of wavelength sources. This behavior is motivated by the higher number of comparably smaller traffic demands of this topology compared to Germany with the used traffic model.

\subsection{Flexible-FSR MWS}
The planning results for different ART values on the Germany topology are shown in Fig.~\ref{fig:flex_study_ger}. The provisioned traffic for the considered \TXOSNR~penalties varies slightly while remaining above the ART (Fig.~\ref{fig:flex_study_ger} \textbf{a)}). Since the demanded data rates are on a continuous scale, the next higher feasible data rate configuration is chosen to fulfill the demand, leading to the variations observed. The number of deployed LPs increases with the \TXOSNR~penalty, as visible from the results in Fig.~\ref{fig:flex_study_ger} \textbf{b)}, because low-rate configurations have to be chosen. As shown in Fig.~\ref{fig:flex_study_ger} \textbf{c)}, UP is observed for high ART > 130~Tbit/s. As expected, the UP is lowest for penalty-free SWSs, and it is highest for the 5~dB penalty MWS scenario. The UP of the 1~dB and the 3~dB \TXOSNR~penalty scenarios are instead comparable in the investigated ART range. At 130~Tbit/s the UP of the 1~dB scenario is marginally higher than the UP of the 3~dB scenario, with less than 0.5~\% variation. This small discrepancy originates from the imperfect RCSA. In this case, as demands are fulfilled sequentially, in the 1~dB scenario initially, more demands can be met than in the 3~dB scenario leading to the spectrum being blocked for a later demand and thereby an overall lower UP. Finally, in Fig.~\ref{fig:flex_study_ger} \textbf{d)}, the distribution of additional percentage of LPs over all ART values are shown. While the 1~dB \TXOSNR~penalty leads to a worst case of 2.2~\% increase, for 3~dB penalty up to 6.8~\% increase is observed and for 5~dB up to 13.6~\%. Furthermore, the variance of the distributions increases with the \TXOSNR~penalty as the effect of the penalty varies with path length.

Fig.~\ref{fig:flex_study_eu} shows the planning results on the EU topology. It can be observed that the differences with varying \TXOSNR penalty are lower than for the Germany topology. This behavior is due to by the lower impact of the \TXOSNR~ on the overall \SNR~for longer paths. From Table~\ref{tab:table-topologies} it can be observed that the average path length on the continental EU topology is 1100~km, representing a $\approx 2.6$-fold increase over the average path length of 420~km in the Germany topology. As a result, the UP is the same for the SWS, the 1~dB and the 3~dB scenario in the considered ART range and the worst case increase in the deployed number of lightpaths lies at 2.2~\%, 3.3~\% and 6.3\% for 1~dB, 3~dB and 5~dB \TXOSNR~penalty, respectively.

\begin{figure}[t]
\centering
\includegraphics[width=\columnwidth]{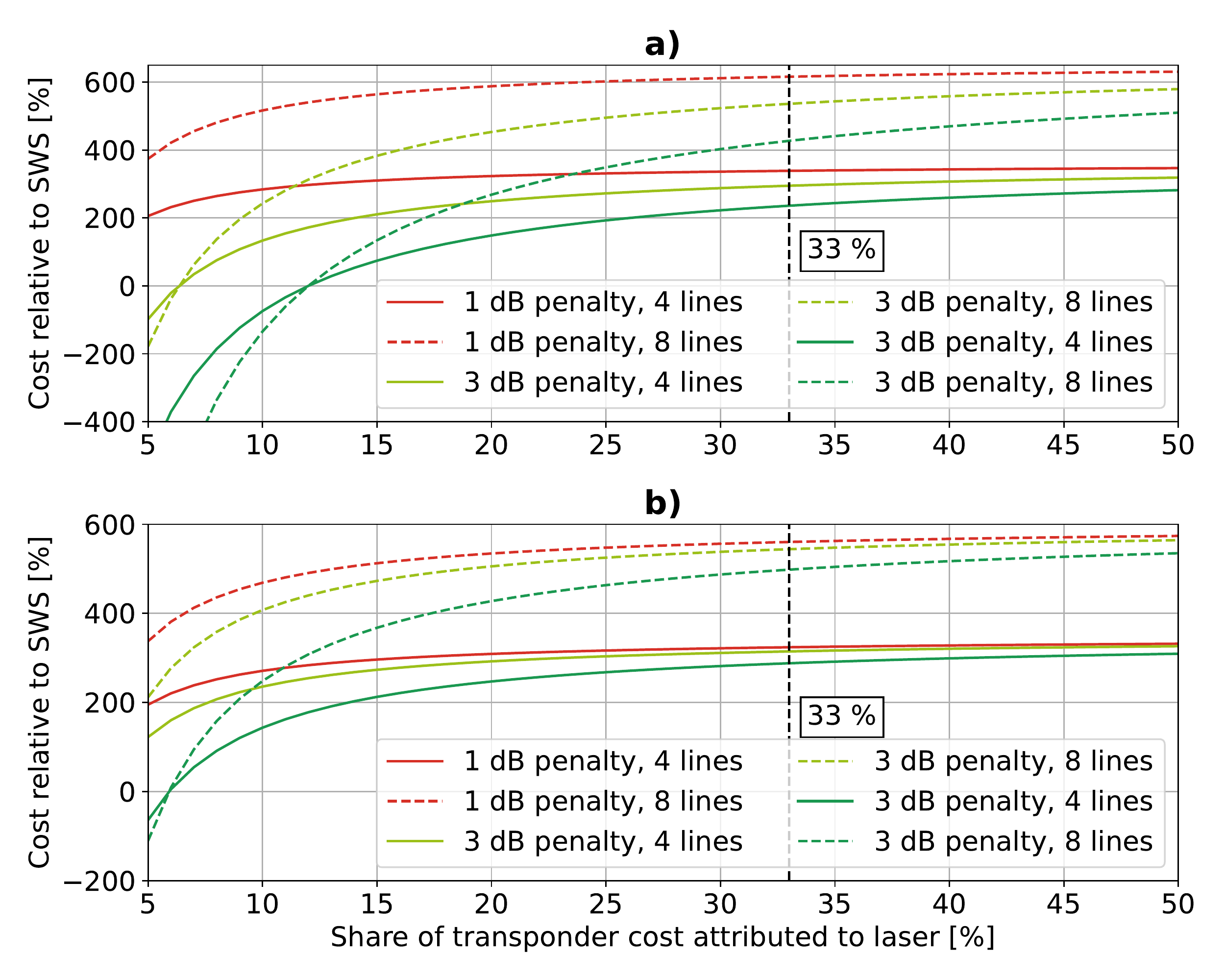}
\caption{\rev{Maximum MWS block cost (including Demux and comb amplifiers) for flexible-FSR MWSs relative to SWS to achieve transponder cost savings versus the share of the laser on the overall transponder cost for Germany \textbf{a)} and EU \textbf{b)} topologies.}}
\label{fig:flex_cost}
\end{figure}

\rev{For flexible-FSR MWSs, the planning does not depend on the number of lines as we assume no limitations in terms of center frequency, configuration and routing of the MWSs lines. However, in practice it may be more difficult to manufacture MWSs with higher number of lines resulting in higher \TXOSNR~penalties. We analyze the potential cost benefits of flexible-FSR MWSs considering the range of 1~dB to 5~dB \TXOSNR~penalty and 4-line as well as 8-line MWSs. For this analysis, we assume that the architecture displayed in Fig.~\ref{fig:CombTxArch} \textbf{c)} is required. Therefore, when compared to the SWS architecture (Fig.~\ref{fig:CombTxArch} \textbf{a)}) an additional Demux as well as one CA per line are required. In the following, we derive the maximum feasible cost of this MWS block in relation to the cost of an SWS for 4-line and 8-line flexible-FSR MWSs.} As explained above, additional LPs have to be deployed  and in some cases an increase in UP is observed. On the other hand, while the number of wavelength sources equals the number of lightpaths in the SWS scenario, significant savings can be achieved in case of 4-line or 8-line flexible-FSR MWSs. As a remark, in addition to the results shown, we computed that the number of required 4-line (8-line) MWSs equals at most 29~\% (17~\%) of the number of LPs in the considered scenarios. 
% Finally, we analyze the potential cost benefits of flexible-FSR MWSs. As explained above, additional LPs have to be deployed  and in some cases an increase in UP is observed. On the other hand, while the number of wavelength sources equals the number of lightpaths in the SWS scenario, significant savings can be achieved in case of 4-line or 8-line flexible-FSR MWSs. As a remark, in addition to the results shown, we computed that the number of required 4-line (8-line) MWSs equals at most 29~\% (17~\%) of the number of LPs in the considered scenarios. 
% It can't simply be divided by 4 (8) as the number of lightpaths at each node is not necessarily a multiple of 4 (8). 
Accounting for these numbers and the additionally placed LPs in each scenario, we estimate the maximum viable cost of a flexible MWS \rev{block} depending on the share of the transponder cost attributed to the laser. The results of this analysis are shown in Fig.~\ref{fig:flex_cost}. On the Germany topology, the variance of the values over \TXOSNR~penalty is higher than on the EU topology as the \TXOSNR~penalty has a higher impact on the network planning. Assuming that the laser has a 33~\% share of the overall transponder cost, depending on the \TXOSNR~penalty, for the Germany topology a 4-line flexible-FSR MWS can cost up to 2.3 to 3.4 times the price of a SWS to enable cost savings. When considering the EU topology, the values lie instead between 2.8 times and 3.2 times. For the 8-line flexible-FSR MWS \rev{block} these numbers lie between 4.2 to 6.1 times for Germany and 5 to 5.6 times for EU. Therefore, our analysis shows, that cost benefits are achieved if the flexible-FSR MWSs \rev{block} can be manufactured at less than 6 times the cost of an SWS. 
% Assuming the architecture shown in Fig.~\ref{fig:CombTxArch} \textbf{c)}, the cost of the flexible-FSR MWS is assumed to include the Demux as well as comb amplifiers for each line.

\subsection{Discussions}
The planning results show that flexible-FSR MWSs should be the preferred solution for mesh core network operators. These MWSs offer the same flexibility as SWSs for the planning of a network. Additional LPs have to be placed to fulfill all demands. Depending on the incurred \TXOSNR~penalty compared to SWSs, the traffic load and the specifics of a network topology, the percentage of additional LPs varies between 2 to 14~\% in our studies.
In return, for the 4-line (8-line) flexible-FSR MWS the number of required wavelength sources equals only 29~\% (17~\%) of the number of SWSs that would be needed for the same LPs. Thereby, the cost of the MWS can be up to 6 times the cost of an SWS to be economically advantageous. 

% While the flexible-FSR MWSs show only minor draw-backs in the network planning results, the widespread deployment of fixed-FSR MWSs causes a severe decrease in requested traffic that can be handled without the underprovisioning of demands. To mitigate this, fixed-FSR MWSs should be only used in addition to SWSs, with the MWSs being deployed only for large demands that need multiple LPs. This reduces the benefits in savings in number of wavelength sources. On the Germany topology, strategies that can fulfill 120~Tbit/s without underprovisioning deploy around 90~\% of the wavelength sources that would be needed when only using SWSs. Therefore, the potential for cost benefits is limited. The exploitation of DSP benefits \cite{DSP-CPE,lundberg2018frequency,DSP-MIMO} offered by fixed-FSR MWSs requiring co-propagation can potentially increase their spectral efficiency while the development of MWS-aware RSAs \cite{dallaglio_impact_2014} generalized to include the configuration selection of LPs can improve the planning in the presence of the additional restrictions to the RCSA. The presented SWS + fixed-FSR MWS hybrid deployment strategies serve as an attempt to a simple solution for this.

%gabriele version below
While the flexible-FSR MWSs show only minor drawbacks in the network planning results, with standard planning strategies the widespread deployment of fixed-FSR MWSs causes a severe decrease in requested traffic that can be handled without underprovisioning. In this context, fixed-FSR MWSs are only sensible as a supplementary solution to SWSs, to be deployed only for large demands that need multiple LPs. On the other hand, this reduces the benefits concerning the savings in the required number of wavelength sources, thus limiting the cost benefits as discussed above for the example of 120~Tbit/s on the Germany topology. Notably, these numbers could be improved through the exploitation of joint DSP benefits that are unique to co-propagating superchannels based on MWSs \cite{DSP-CPE,lundberg2018frequency,DSP-MIMO} and through the development of MWS-aware RSAs \cite{dallaglio_impact_2014}. The presented SWS + fixed-FSR MWS hybrid deployment strategy serves to prove that straightforward applicability of fixed-FSR MWSs in current network configurations is challenging. \rev{Additionally, a hybrid strategy leads to more equipment types deployed in the network, further complicating its operation.} The development of optimal planning solutions and the assessment of the ultimate capacity of multi-wavelength transponders in a networking context are instead regarded as future research.

% \section{Conclusions}
% In this work, specifications of state-of-the-art MWSs for high-bandwidth transponder configurations are developed in Section \ref{sec:architectures}. We compare 4-line and 8-line fixed-FSR MWSs and flexible-FSR MWSs with the SWS baseline in network planning studies. For fixed-FSR MWSs we show that only hybrid SWS and fixed-FSR MWS deployment strategies are sensible in mesh core network topologies. For a throughput of 120~Tbit/s, a limited reduction in the number of wavelength sources to 90~\% of the number of LPs is achieved on the Germany topology at the expense of network capacity leading to higher underprovisioning when compared to SWSs. For flexible MWSs we show savings in the number of wavelength sources of around 71~\% for 4-line MWSs and 83~\% for 8-line MWSs as compared to the use of SWSs. In exchange, depending on the \TXOSNR~penalty, up to 14\% additional lightpaths are required to fulfill the demands on the Germany topology while an increase of up to 6~\% is needed for EU. Therefore, depending on the network topology and the exact share of the laser in overall transponder cost, an 8-line flexible-FSR MWS can be economically feasible at up to 6 times the cost of an SWS. In our assessment, the use of fixed-FSR MWSs in mesh core networks is limited and situational, while the use of flexible-FSR MWSs can lead to in significant cost savings and efficiency improvements in these networks.

\section{Conclusions}
In this work, specifications of state-of-the-art MWSs for high-bandwidth transponder configurations are developed (Sec.~\ref{sec:architectures}). We compare 4-line and 8-line fixed-FSR MWSs and flexible-FSR MWSs with the SWS baseline in network planning studies. We show that for fixed-FSR MWSs only hybrid deployment strategies employing both SWS and fixed-FSR MWS are sensible in mesh core network topologies. The potential cost reduction benefits are limited in the fixed-FSR scenario as more complex network planning algorithms and lightpath re-configuration strategies are required to overcome the restrictions that fixed-FSR MWSs impose on the planning. For flexible MWSs, we show that significant savings in the number of wavelength sources of around 71~\% for 4-line MWSs and 83~\% for 8-line MWSs are achievable. We observe that this large reduction comes at the cost of only a moderate increase in the required number of lightpaths, up to 14\% depending on the \TXOSNR~penalty and the network topology considered. As a result, flexible-FSR MWS show the potential for significant cost- and power-efficiency improvements as we observe that they can be economically feasible at up to 6 times the cost of an SWS. Therefore, it is our assessment that for mesh core network applications the focus in MWS development should lie in flexible-FSR MWSs offering high \TXOSNR.

\section*{Funding}The work has been partially funded by the German Federal Ministry of Education and Research in the project STARFALL (16KIS1418K) and the programme of "Souverän. Digital. Vernetzt." joint 
project 6G-life (16KISK002), as well as the European Research Council through the ERC-CoG FRECOM project (grant no. 771878).

% \section*{Acknowledgments}The authors thank H. Haase, C. Wiede, and J. Gabler for technical support.

% \section{References}

% Full references (to aid the editor and reviewers) must be included. This will be produced automatically if you are using a .bib file.

% \bigskip
% \noindent Add citations manually or use BibTeX. See \cite{Chitimalla:17,Wen:16}.

% Bibliography
\bibliography{references}

\end{document}